\newcount\style \style=1
\ifodd\style
  \documentstyle[aas2pp4,psfig]{article}
\else
  \documentstyle[12pt,aasms4]{article}
\fi

\begin{document}

\def\lmax{l_{\rm max}}
\def\h1{ \ion{H}{1} }
\def\spose#1{\hbox to 0pt{#1\hss}}
\def\ltsim{\mathrel{\spose{\lower.5ex\hbox{$\mathchar"218$}}
     \raise.4ex\hbox{$\mathchar"13C$}}}
\def\gtsim{\mathrel{\spose{\lower.5ex \hbox{$\mathchar"218$}}
     \raise.4ex\hbox{$\mathchar"13E$}}}

\title{Warped Galaxies From Misaligned Angular Momenta}

\author{Victor P.\ Debattista\altaffilmark1 and J.\ A.\ Sellwood}
\affil{Department of Physics and Astronomy, Rutgers, The State University of New 
Jersey, 136 Frelinghuysen Road, Piscataway, NJ 08854-8019}
\affil{debattis@astro.unibas.ch, sellwood@physics.rutgers.edu}
\altaffiltext{1}{Current address: Astronomisches Institut, Universit\"at Basel, 
Venusstrasse 7, CH-4102 Binningen, Switzerland}

\begin{abstract}
A galaxy disk embedded in a rotating halo experiences a dynamical friction force 
which causes it to warp when the angular momentum axes of the disk and halo are 
misaligned.  Our fully self-consistent simulations of this process induce 
long-lived warps in the disk which mimic Briggs's rules of warp behavior.  They 
also demonstrate that random motion within the disk adds significantly to its 
stiffness.  Moreover, warps generated in this way have no winding problem and 
are more pronounced in the extended \h1\ disk.  As emphasized by Binney and his 
co-workers, angular momentum misalignments, which are expected in hierarchical 
models of galaxy formation, can account for the high fraction of warped 
galaxies.  Our simulations exemplify the role of misaligned spins in warp 
formation even when the halo density is not significantly flattened.

\end{abstract}

\section{Introduction}
An ``integral sign'' twist has been observed in the extended \h1\ disk of many 
galaxies; in some cases it can also be seen in the star light.  Briggs (1990) 
characterized the behavior of a sample of 12 warped galaxies as: (1) coplanar 
inside $R_{25}$, and warped beyond, with a straight line of nodes (LON) inside 
$R_{\rm Ho}$, (2) changing near $R_{\rm Ho}$, (3) into a LON on a leading spiral 
(as seen from the inner disk) outside $R_{\rm Ho}$.  Bosma (1991) found 12 
clearly warped disks in a sample of 20 edge-on systems; taking into account 
random warp orientation, the true fraction of warped disks must be larger.  This 
high fraction of warped galaxies implies either that warps are long lived 
features or that they are repeatedly regenerated.

If a twisted disk were modeled as a set of uncoupled tilted rings in a 
flattened potential, their changing precession rates with radius would lead
to a winding problem, similar to that for spirals (e.g.\ Binney \& Tremaine 
1987).  If warps are long-lived, therefore, some means to overcome differential 
precession is required (see reviews by Toomre 1983 and Binney 1992).

Most recent ideas for warp formation rely in some way on the influence of the 
halo.  Toomre (1983) and Dekel \& Shlosman (1983) suggested that a flattened 
halo misaligned with the disk can give rise to a warp, and Toomre (1983), Sparke 
\& Casertano (1988) and Kuijken (1991) found realistic warp modes inside rigid 
halos of this form.  However, angular momentum conservation requires there to 
be a back reaction on the halo (Toomre 1983; Binney 1992); Dubinski \& 
Kuijken (1995) and Nelson \& Tremaine (1995) showed that a mobile halo should 
cause a warped disk to flatten quickly (but see also Binney et al.\ 1998).

As a warp represents a misalignment of the disk's inner and outer angular 
momenta, Ostriker \& Binney (1989) proposed a qualitative model in which the 
warp is generated by the slewing of the galactic potential through accretion 
of material with misaligned spin.  The accretion of satellites by larger 
galaxies, such as in the Milky Way, provides direct evidence of late-infalling 
material with misaligned angular momentum.  Such misalignments are expected to 
be generic in hierarchical models of galaxy formation (Quinn \& Binney 1992) 
in which the spin axis of late-arriving material, both clumpy and diffuse, is 
poorly correlated with that of the material which collapsed earlier.  Jiang \& 
Binney (1998) calculate a concrete example of a warp formed through the 
addition of a misaligned torus of matter.

If the accreted material is flattened due to its intrinsic spin, a density 
misalignment is present immediately which exerts a torque to twist the disk, as 
in Jiang \& Binney's model.  But dissipationless halo material may not be 
strongly flattened despite streaming about a misaligned axis.  Nevertheless such 
material will exert a torque on the disk through dynamical friction, causing the 
disk's angular momentum vector to tip towards alignment with that of the halo, 
as we show in this {\it Letter}.

Rotation in the inner halo will cause the disk to tilt differentially because 
the inner disk experiences a stronger dynamical friction torque (since the 
densities of both disk and halo are highest) and because time-scales are 
shortest in the center.  As the inner disk begins to tip, the usual 
gravitational and pressure stresses in a twisted disk become important.  Our 
fully self-consistent $N$-body simulations show that this idea is actually quite 
promising and leads to warps which are relatively long-lived and of the observed 
form.

Dynamical friction arises through an aspherical density response in the halo 
that lags ``downstream'' from the disk.  In the long run, however, the density 
distribution in the halo must become symmetric about the disk plane, even while 
it continues to rotate about a misaligned axis.  Simple time-reversibility 
arguments dictate that a steady system cannot support net torques (cf.\ the 
``anti-spiral'' theorem, Lynden-Bell \& Ostriker 1967; Kalnajs 1971).  We 
therefore find the dynamical friction torque on the disk does not persist 
indefinitely.  We do not regard this as a serious objection to our model, since 
late infalling material must constantly revise the net angular momentum of the 
halo.

\section{Numerical Method}
Since mild force anisotropies in many grid-based $N$-body methods can cause an 
isolated disk to settle to a preferred plane (e.g.\ May \& James 1984), we 
adopt a code with no preferred plane.  An expansion of the potential in 
spherical harmonics has been widely used for $N$-body simulations both with a
grid (van Albada 1982) and without (Villumsen 1982; White 1983; McGlynn 1984). 
Here we adopt an intermediate approach: we tabulate 
coefficients of a spherical harmonic expansion of the density distribution on a 
radial grid, and interpolate for the gravitational forces between the values on 
these shells.  The radial grid smooths the gravitational field, thereby avoiding 
the problem of ``shell crossings.''  Since there is no gridding in the angular 
directions, we retain the full angular resolution up to the adopted $\lmax$ -- 
the maximum order of the spherical harmonic expansion.

While avoiding a preferred plane, this method is not well-suited to 
representation of disks.  The vertical restoring force to the disk mid-plane 
converges slowly with increasing $\lmax$, as shown in Figure 1.  Most of our 
simulations included terms to $\lmax = 10$ only; tests with higher $\lmax$ (and 
fewer particles) suggest these models overestimate the magnitude and duration of 
the warp in the massive part of the disk, although milder and shorter-lived 
warps still develop.  Moreover, the warp in the test-particle layer beyond the 
edge of the massive disk is unaffected by force resolution.

\ifodd\style
  \begin{figure}[t]
  \centerline{\psfig{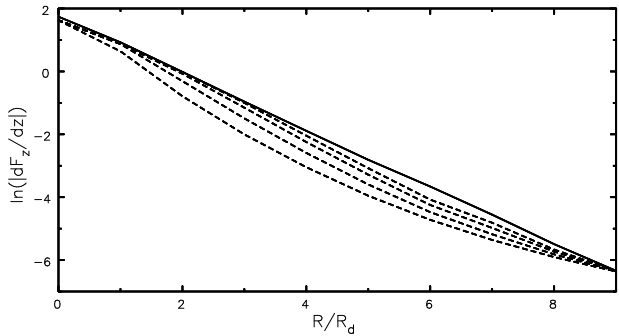}}
  \caption{The mid-plane force gradient for the disk described in section 4.
    The solid line is the exact gradient, the dashed lines show values obtained
    with $l_{\rm max} = 10$, 20, 35 \& 50 in order of decreasing error.}
  \end{figure}
\fi

\section{Initial model}
While we have found that a massive halo is able to produce spectacular warping, 
we here prefer more realistic minimal halo models (cf.\ Debattista \& Sellwood 
1998).  Our galaxy model has three massive components: an exponential disk of 
length-scale $R_d$, truncated at $8 \ R_d$, a polytropic halo and a central 
softened point mass.  The ratio disk:halo:central mass is 1:9:0.2, chosen to 
give a roughly flat rotation curve out to $\sim 15 \ R_d$.  We also include a 
disk of test particles, extending well outside the massive disk, that is 
intended to mimic the neutral hydrogen layer of a galaxy.

The initial massive disk had velocity dispersions set by adopting Toomre $Q = 
1.5$ and a thickness of $0.1 \ R_d$ which were both independent of radius.  The 
massless particles started with exactly circular orbits in the disk mid-plane.  
The central mass is a single particle with a core radius of $0.3 R_d$.

The initial distribution of halo particles was generated by the method first 
used by Raha et al.\ (1991) which gives an exact equilibrium isotropic halo in 
the potential of the disk and central mass.  The halo extends to a radius 
$r_{\rm trunc} = 30~R_d$.  Because the disk has mass, the halo is not precisely 
spherical; its initial axis ratio varies from closely spherical at $r \geq 4 
R_d$ to $c/a \simeq 0.7$ near $r = 1.5~R_d$.  The initial halo angular momentum 
was created by selectively reversing halo particle velocities about a chosen 
axis, which for Run 1 is tipped away from the disk spin ($z$-)axis by
$45^\circ$ in the $x$-direction.  We chose a value of the dimensionless
$\lambda \equiv \frac{L}{G} \sqrt{\frac{\mid E \mid}{M^5}} 
\simeq 0.07$ for our halo models; here $L$, $E$ and $M$ are respectively the 
total angular momentum, energy and mass of the halo.  A value of 0.07 is typical 
in hierarchical clustering models (Barnes \& Efstathiou 1987; Steinmetz \& 
Bartelmann 1995).  

We work in units where $G = M\ ( = M_{\rm disk} + M_{\rm halo}) = R_d = 1$; the 
unit of time is therefore $(R_d^3 / GM)^{1/2}$.  A rotation period in the disk 
plane at $R=1$ is 32.  We set $\lmax=10$, used a radial grid with 200 shells and 
a time step of 0.05.  The disk and halo components are represented by a total of 
$10^6$ equal mass particles.  Our simulations conserve energy to better than 
$0.04\%$.

\ifodd\style
  \begin{figure}[t]
  \centerline{\psfig{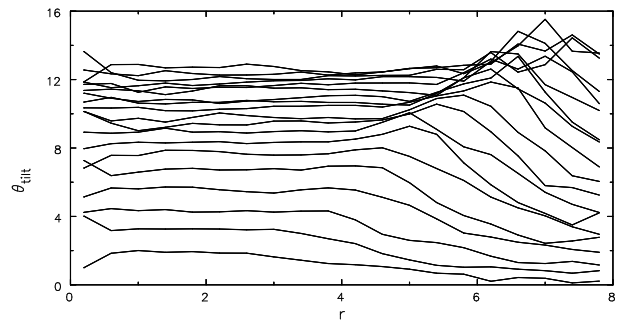}}
  \caption{The disk tilt angle (in degrees) as a function of radius at 
    intervals of 50 time units ($t=50$ to $t=950$ from bottom up).}
  \end{figure}
\fi

\ifodd\style
  \begin{figure}[t]
  \centerline{\psfig{figure=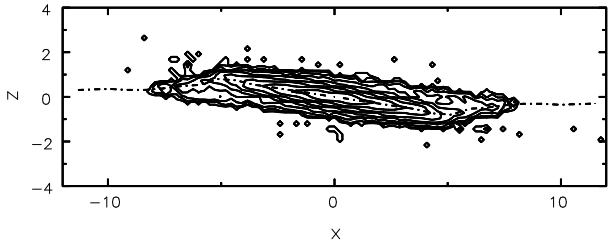,width=1.0\hsize,clip=,angle=0}}
  \caption{Contours of the disk density in Run 1 at $t = 400$ projected onto
    the $(x,z)$ plane.  The dot-dash line indicates the cross-section of the
    layer of massless particles.  By this time, the inner disk has tilted
    $\sim 10^\circ$ away from its original plane, which was horizontal.}
  \end{figure}
\fi

\section{Results}

\subsection{Warp in the massive disk}
The outer disk lags as the inner part of the disk in Run 1 begins to tilt, 
causing a warp to develop almost at once.  Figure 2 shows that the approximately 
rigid tilt of the inner disk increases rapidly at first and then more slowly, 
while the radius at which the warp starts also moves outwards over time.  The 
disk in Run 1 at $t=400$ is shown in Figure 3.

\ifodd\style
  \begin{figure}[t]
  \centerline{\psfig{figure=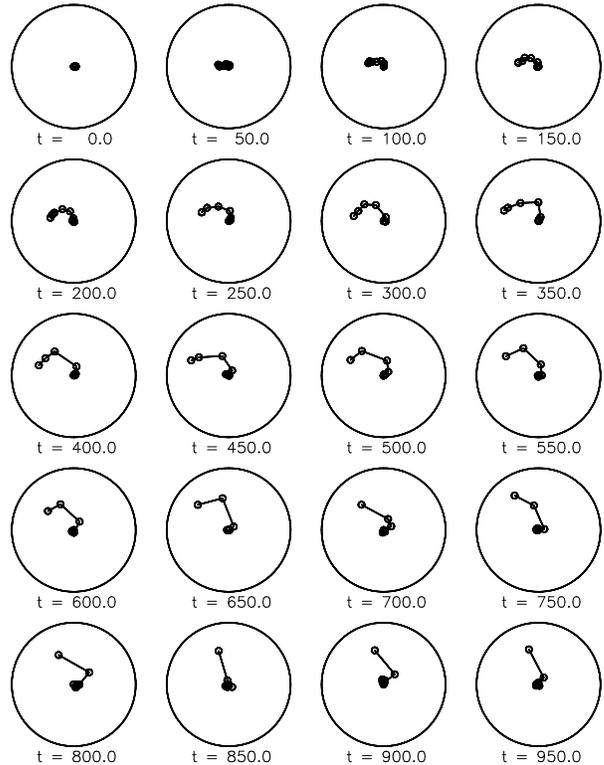,width=1.0\hsize,clip=,angle=0}}
  \caption{Polar angle plots showing the shape of the warped disk.  The
    points indicate the angles $(\theta,\phi)$ of the symmetry axis of
    an annular piece of the disk (width $0.8~R_d$) relative to the inner
    disk axis at the given time.  The radial coordinate is the polar angle
    $\theta$ and the azimuthal coordinate shows the azimuth, $\phi$, of the 
    LON of each annulus.  The boundary circle is at $\theta = 10^\circ$.  Disk 
    rotation is counter-clockwise.}
  \end{figure}
\fi

Figure 4 shows the warp of the massive disk in the form of a Tip-LON diagram 
(Briggs 1990) at equally spaced times.  Each point represents the direction of 
the normal to the best-fit plane of an annular piece of the disk, with the 
center of the disk defining the origin.  The normal to the inner disk tilts 
initially in the $(x,z)$ plane while the outer disk is left behind, thereby 
shifting outer-disk points along the $\phi = 180^\circ$ direction (e.g.\
$t=100$) while the almost flat inner part of the disk gives rise to the 
concentration of points in the center.  The warp reaches a maximum angle of $\sim 7^\circ$ at $t \simeq 350$ and it takes roughly 700 time units ($\sim 20$ disk rotations at $R=R_d$) for most of the disk to settle back to a common plane.

The leading spiral, reminiscent of Briggs' (1990) third rule, develops through clockwise differential precession in the outer parts.  Precession is a consequence of gravitational coupling between the inner and outer disk (Hunter \& Toomre 1969).  The extremely slow precession of the outermost edge of the disk indicates that it is subject to a very mild torque, arising almost exclusively from the distant, tipped inner disk.  The outer disk would precess more rapidly if the halo density distribution were flattened.

\subsection{Secular evolution}
Several changes occur as the model evolves:  the dynamical friction force 
driving the warp decays over time, as expected from \S1, causing the inner 
disk to tilt more slowly (Figure 2).  The radius at which the warp starts 
moves outwards and the amplitude of the warp (the difference in tilt angle 
between the inner and outer disk) also decreases.  The massive disk is almost 
coplanar again by time 1000 (Figure 4).  Spiral arms and a weak bar also 
drive up the velocity dispersion of the particles to $Q \simeq 2.5$.

The inner disk tilts remarkably rigidly indicating strong cohesion which arises 
from two distinct mechanisms.  Most studies have 
focused on gravitational forces, which Hunter \& Toomre (1969) found were 
inadequate to persuade the outer disk to precess along with the inner disk in a 
steady mode.  However, the disk is also stiffened by the radial epicyclic 
excursions of the stars which communicate stresses across the disk.

Both self-gravity is strongest and random motion is greatest in the inner disk, 
where the coupling is evidently strong enough to preserve its flatness.  The 
settling of the disk to ever larger radii should be describable in terms of the 
group velocity and/or damping of bending waves in a warm and finitely thick 
disk, but the absence of a dispersion relation valid in this regime precludes 
a comparison with theory.  
It is interesting that each annulus settles as its precession angle reaches $\sim180^\circ$ (Figure 4), thereby preventing excessive winding of the warp.  
The settling of each ring after half its precession period could be coincidental 
but we have seen it in many models.  One possible reason is that a warm disk 
cannot support bending waves with wavelength shorter than the average epicycle 
diameter; a prediction based on this idea is only roughly in accord with the 
radially dependent settling time, however.  It should be noted that whether 
settling is described by group velocity, wave damping or precession angles, it 
should be more rapid in a disk with stronger forces towards the mid-plane.

To demonstrate the importance of random motion, we ran a new simulation (Run 
2) identical to Run 1 except with $Q = 4.0$ initially in the disk.  The warp was 
much reduced, as shown in Figure 5, even though the inner disk tilted by an 
angle comparable to that in Run 1. 

\ifodd\style
  \begin{figure}[t]
  \centerline{\psfig{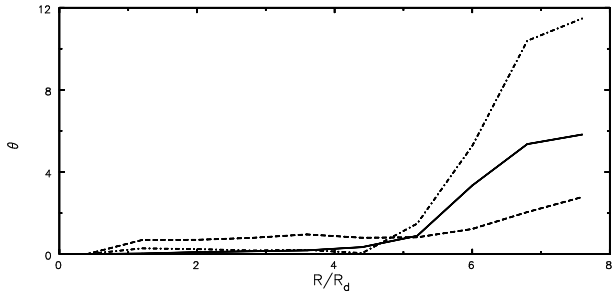}}
  \caption{The warp angle as a function of radius for three different models.
    Solid line: our standard (run 1), dashed line: hotter disk (run 2) and
    dot-dashed line: greater misalignment (run 3).  All warp angles are at 
    $t = 450$.}
  \end{figure}
\fi

\subsection{Test particle layer}
The sheet of test particles is intended to approximate a gaseous disk.  Being 
massless, it does not induce a response from the halo, but is perturbed by 
forces from the tilted massive disk and its associated halo response.  Within 
$8 \ R_d$ the test particles simply tilt or warp with the disk.  Outside the 
massive disk, however, the 
disturbance forces from the halo and the massive disk drop off rapidly and the 
plane of this dynamically-cool sheet hardly moves at large radii.  It therefore 
appears warped relative to the plane of the tilted disk, and remains so even by 
the end of the simulation when the stellar disk has mostly settled.

\subsection{Further Simulation}
The spin axis of the halo in Run 1 was initially inclined at 45$^\circ$ to 
that of the disk.  In Run 3, we set this angle to 135$^\circ$ thereby reversing
the sign of $J_z$, the component of the halo's angular momentum along the 
$z$-axis.  The larger misalignment angle causes the 
inner disk to tilt faster and further and gives rise to a larger warp, as shown 
in Figure 5.  Dynamical friction also lasts longer.  Other properties 
of the warp are similar to those in Run 1; in particular, the warp begins at a 
similar radius at equal times.

\section{Conclusions}
Our simulations have confirmed that dynamical friction from a halo having 
angular momentum misaligned with that of the disk causes a transient warp.  The 
warp has two properties commonly observed: the LON traces out a leading spiral 
relative to the inner disk and lasts longest in the \h1\ layer.

By driving the warp, we side-step the most troublesome difficulties faced by 
other warp mechanisms.  The bane of global mode warp models, that forces are 
simply too weak to overcome differential precession near the edge, has become a 
strength in our mechanism: the weak coupling of the outer edge {\it creates\/} 
the warp.  Furthermore, the gradual settling of the warm disk avoids any winding 
problem.

The massive disk can warp slightly, but is largely rigid both because of 
gravitational restoring forces and radial pressure.  The size and lifetime of 
the warp in the massive disk are probably somewhat overestimated because our
numerical method does not yield the full gravitational restoring force.  This 
worry does not affect the conclusions about the warp in the extended \h1\ 
layer, which has very little mass and rigidity.

We have deliberately adopted an almost spherical halo in order to show that 
warps can be formed without misaligned density distributions.  Rotating halos 
are likely, of course, to be slightly flattened also, in which case the disk 
will respond to both types of forcing.  This will lead to warps that precess, 
\h1\ layers that do experience forces, and so on.  Studies of these cases will 
be reported in a future paper.

As noted above, we expect the net angular momentum of the halo to be revised as 
material continues to straggle in long after the main galaxy has reached 
maturity.  Every change to the halo's spin vector can be expected to affect the 
disk through friction, even if the arriving material is torn apart at large 
distance by the tidal field of the host galaxy.  Our picture is similar to that 
proposed by Ostriker \& Binney (1989), but who envisage warps as being driven 
from the outside by a misalignment of the inner disk with the flattened outer 
halo.  In practice, both mechanisms must be inextricably linked.  On-going 
infall makes it hardly surprising that warps are detected in most disks.

\bigskip
Conversations with Scott Tremaine and Alar Toomre and the report of the referee, 
James Binney, were most helpful.  The authors wish to thank the Isaac Newton 
Institute, Cambridge, England for their hospitality for part of this project.  
This work was supported by NSF grant AST 96/17088 and NASA LTSA grant NAG 
5-6037.

\newpage

\ifodd\style
\else
\centerline{\bf Figure Captions}

\bigskip
Fig. 1. {The mid-plane force gradient for the disk described in section 4.
    The solid line is the exact gradient, the dashed lines show values obtained
    with $l_{\rm max} = 10$, 20, 35 \& 50 in order of decreasing error.}

Fig. 2. {The disk tilt angle (in degrees) as a function of radius at 
    intervals of 50 time units ($t=50$ to $t=950$ from bottom up).}

Fig. 3. {Contours of the disk density in Run 1 at $t = 400$ projected onto
    the $(x,z)$ plane.  The dot-dash line indicates the cross-section of the
    layer of massless particles.  By this time, the inner disk has tilted
    $\sim 10^\circ$ away from its original plane, which was horizontal.}

Fig. 4. {Polar angle plots showing the shape of the warped disk.  The
    points indicate the angles $(\theta,\phi)$ of the symmetry axis of
    an annular piece of the disk (width $0.8~R_d$) relative to the inner
    disk axis at the given time.  The radial coordinate is the polar angle
    $\theta$ and the azimuthal coordinate shows the azimuth, $\phi$, of the 
    LON of each annulus.  The boundary circle is at $\theta = 10^\circ$.  Disk 
    rotation is counter-clockwise.}

Fig. 5. {The warp angle as a function of radius for three different models.
    Solid line: our standard (run 1), dashed line: hotter disk (run 2) and
    dot-dashed line: greater misalignment (run 3).  All warp angles are at 
    $t = 450$.}

Fig. 6. {The warp angle, at $t = 100$, as a function of radius for 
    $\lmax = 10$ (solid line), 16 (short dashed line), 20 (dot-dashed line), 
    25 (dotted line) and 35 (long dashed line).}
  
\fi

\newpage

\end{document}